\documentclass[aps,pra,twocolumn,groupedaddress,amsmath,amssymb]{revtex4}

\usepackage{epsfig}
\usepackage{amssymb}
\usepackage{amsmath}
\usepackage{graphicx}
\usepackage{amsfonts}
\usepackage{epsfig}
\usepackage{revsymb}
\usepackage{bm}
\usepackage{amsmath}
\usepackage{amssymb}
\usepackage{latexsym}
\usepackage{thumbpdf}
\usepackage{hyperref}

\begin{document}

\title{Nucleation at quantized vortices and the heterogeneous phase separation in supersaturated superfluid $^3$He-$^4$He liquid mixtures}

\author{S.N. Burmistrov and L.B. Dubovskii}
%\email[]{burmistrov_sn@nrcki.ru}
\affiliation{NRC "Kurchatov Institute", 123182 Moscow, Russia 
\\
and  
\\
Moscow Institute of Physics and Technology, 141700 Dolgoprudnyi, Russia
\\
E-mail: burmistrov\_sn@nrcki.ru }

%\date{today}

\begin{abstract}
Supersaturated superfluid $^3$He-$^4$He liquid mixture, separating into the $^3$He-concentrated \textit{c}-phase and  $^3$He-diluted  \textit{d}-phase, represents a unique possibility for studying macroscopic quantum nucleation and quantum phase-separation kinetics  in binary mixtures at low temperatures down to absolute zero. One of possible heterogeneous mechanisms for the phase separation of supersaturated \textit{d}-phase is associated with superfluidity of this phase and with a possible existence of quantized vortices playing a role of nucleation sites for the \textit{c}-phase of liquid mixture. We analyze the growth dynamics of vortex core filled with the \textit{c}-phase and determine the temperature behavior of \textit{c}-phase  nucleation rate and the crossover temperature between the classical and quantum nucleation mechanisms.  
\\
\\

\par
\hspace{3em} 64.60.Q-  Nucleation;
\par
\hspace{3em} 64.70.Ja  Liquid-liquid transitions;
\end{abstract}

\maketitle

\subsection{Introduction}
\par
This year E.Ya. Rudavskii celebrates 80. Our meeting with Eduard Yakovlevich has taken place about the same time when the Department of Physics of Quantum Fluids and Crystals has started the systematic experimental study on the phase separation kinetics of supersaturated $^3$He-$^4$He mixtures. This study has laid the foundations for new field of physics, namely,  macroscopic quantum nucleation or kinetics of first-type phase transitions in condensed matter at temperatures so close to absolute zero that the classical thermal-activation phase-transition mechanism becomes completely ineffective. Under the influence of pioneer experiments and personal charm of Eduard Yakovlevich we, keen at that time with the theory of macroscopic quantum tunneling and the role of dissipative processes,  have turned to the study of the low-temperature phase-separation  kinetics of liquid $^3$He-$^4$He mixtures and the energy dissipation effects associated mainly with the diffusion of impurity $^3$He atoms. 
\par 
In 1969 during the study of degenerated $^3$He-$^4$He liquid mixtures there is demonstrated a possibility of preparing the metastable state of supersaturated superfluid $^3$He-$^4$He liquid mixture in the lack of free liquid-vapor 
surface~\cite{Landau69}.  For $T<$70 mK, there are obtained the long-lived supersaturated liquid mixtures staying in the metastable state for two and more hours. The experimental studies of phase separation kinetics in liquid $^3$He-$^4$He mixtures have been started by Brubaker and Moldover~\cite{Brubaker72} and then continued in 80-ies~\cite{Hoffer80, Alpern82,Bodensohn89}. The experiments were performed in the high temperature $T>$0.5 K region and the main attention was paid for the phase separation in the vicinity of the tricritical point in the connection with the type-II phase transition fluctuation theory developing rapidly those years. The experimental observations have adequately been described within the framework of the classical thermal-activation nucleation theory. 
\par 
Later in the early 90-ies the study is started of the phase separation kinetics of superfluid $^3$He-$^4$He liquid mixtures in the region of lower temperatures $T<$200 mK with the aim of detecting the transition from the thermal-activation phase-separation regime to the quantum underbarrier one. There have been used new methods of experimental study. One method, developed in Kharkov in B.Verkin Institute for Low Temperature Physics and Engineering, is based on a continuous change in the $^3$He concentration directly in the course of experiment at constant pressure and temperature due to varying the osmotic pressure and fountain pressure. The variation rate in the $^3$He concentration is usually about 10$^{-4}$\% per second. The state of the liquid mixture is recorded by two methods~\cite{FNT91,FNT92,FNT92a,FNT94, Edik95}. The acoustic method is based on the sound velocity variation in the course of phase separation of liquid mixture and the capacitive one is based on the measurement of dielectric permittivity. The jump-like reduction in the $^3$He concentration, recorded simultaneously with the  sound and capacitive methods, corresponds to  separating a metastable supersaturated liquid mixture into two phases.
\par 
As is known,  due to existence of the effects of osmotic pressure and fountaining  in a superfluid phase of liquid $^3$He-$^4$He mixture the temperature inhomogeneity leads readily to the pressure and concentration inhomogeneity. To avoid the effect of this factor, the process of preparing a supersaturated liquid mixture should occur under  invariable temperature and sufficiently slowly in order not to set the $^3$He atoms in motion. Thus, employing the dependence of the separation line on the
pressure, the authors of  work \cite{Satoh92} have proposed another scheme of continuous pressure variation using a thin capillary (superleak) through which the superfluid $^4$He component alone can flow. This results in changing the pressure and $^3$He concentration in the experimental cell. In the course of experiment the total $^3$He amount in the cell remains unvaried.  The rate of pressure variation is about 0.25 atm/hr, corresponding to the $^4$He flow as about a microgram per second  and as a concentration rate 6$\cdot$10$^{-6}$\% per sec. This is by a factor of 10$^4$--10$^5$ as compared with that in the phase separation experiments on liquid mixtures in the vicinity of the tricritical point. The final results of the research and their discussion are given in work \cite{Satoh02}. 
\par 
So far, the theoretical interest \cite{LPK,BDT} has mainly been focused on the homogeneous mechanism of phase separation in the supersaturated \textit{d}-phase of liquid $^3$He-$^4$He mixture. In spite of careful analysis a scepticism has remained. The possible mechanisms for inhomogeneous phase separation of liquid mixture have not received a proper attention. In particular, this concerns a possibility of the presence of remnant quantized vortices and an estimate of the crossover  temperature for the thermal and quantum regimes of phase separation. 
\par 
Here we consider one of possible heterogeneous mechanisms for the phase separation in the supersaturated \textit{d}-phase of liquid $^3$He-$^4$He mixture, which is associated with superfluidity of \textit{d}-phase and with possibility of existence of quantized vortices. The vortices, in its turn, can play a role of nucleation sites for the \textit{c}-phase of liquid mixture. The idea  goes directly back to works \cite{Reut,Ohmi,Senbetu} where the phenomenon was analyzed of $^3$He atom adsorption onto the quantized vortex core in superfluid He-II. Unlike the previous works \cite{Jezek95} dealing with the thermal-activation nucleation mechanism alone, we here concentrate our attention on the growth dynamics of vortex core filled with the \textit{c}-phase and determine the nucleation rate in the quantum region as well as the thermal-quantum crossover temperature.  

\subsection{The quantized vortex structure in the saturated $^3$He-$^4$He liquid mixture}

\par 
Let us consider the structure, core radius and energy of rectilinear vortex with one circulation quantum in the saturated superfluid \textit{d}-phase of liquid $^3$He-$^4$He mixture. It is well known that $^3$He atoms tend to be localized and trapped with the vortex core. Below we analyze  a simple model with the rigid core for quantized vortex in the saturated superfluid $^3$He-$^4$He liquid mixture, assuming the vortex core to be filled with the \textit{c}-phase. Beyond the vortex core of radius $R$ the superfluid velocity $V_s$ at distance $r$ from vortex line is subjected to equation:  
$V_s(r)=\hbar /m_4 r$ as a result of circulation quantization $\oint V_s\, dl=2\pi\hbar/m_4$. 
\par
The condition for equilibrium of \textit{d}-phase in the bulk requires the constancy of thermodynamic potentials and temperature 
\begin{gather*}
\Phi (P,\, T,\, Z,\, V_s)+\frac{V_s^2}{2}  =\text{const}  \equiv  \Phi (P,\, T,\, Z)_{|_{r=\infty}},
\\ Z(P,\, T,\, c,\, V_s)  =\text{const} \equiv  Z (P,\, T,\, c)_{|_{r=\infty}},
\\
T  =\text{const}  \equiv  T_{|_{r=\infty}}.
\end{gather*}
Here  $\Phi =\mu _4/m_4$, potential $Z=\mu _3/m_3 -\mu _4/m_4$ is conjugated to the mass $^3$He concentration, and 
 $\mu _{3,\, 4}$ and $m_{3,\, 4}$ are the chemical potentials and masses of  $^3$He and $^4$He atoms, respectively.  
\par
Using the known variations for the thermodynamic potentials 
\begin{gather*}
\delta\Phi  =  \frac{\delta P}{\rho} -\sigma\,\delta T -c\,\delta Z-\,\frac{\rho
_n}{\rho}\,\delta\!\left(\frac{V_s^2}{2}\right),
\\
\delta Z  =  \frac{\partial}{\partial c}\left(\frac{1}{\rho}\right)\,\delta P
-\,\frac{\partial\sigma}{\partial c}\,\delta T +\,\frac{\partial Z}{\partial c}\,\delta c -\,\frac{\rho
_n}{\rho}\,\delta\!\left(\frac{V_s^2}{2}\right), 
\end{gather*}
we find that the variations of concentration  $c$ and pressure $P$ are determined with the following equations 
\begin{eqnarray}\label{f904}
\frac{\partial Z}{\partial c}\,\nabla c  =  -\,\frac{1}{\rho}\,\frac{\partial\rho _s}{\partial
c}\,\nabla\frac{V_s^2}{2}, \;\;\;\;\;
\nabla P  =  -\,\rho _s\nabla\frac{V_s^2}{2}.
\end{eqnarray}
\par
Let us put $P$, $c$ and $T$ for the magnitudes of pressure, concentration and temperature far from the vortex at the infinity 
$r=\infty$. Next, while the pressure and concentration variations are small, i.e.  $\delta P\ll P$ and $\delta c\ll c$, we find approximately 
\begin{eqnarray}\label{f905}
P(r)-P & \approx & -\,\rho _s\,\frac{V_s^2}{2},
\\
c(r)-c & \approx & -\,\frac{\partial\rho _s/\partial c}{\rho\,\partial Z/\partial c}\,\frac{V_s^2}{2}. \nonumber
\end{eqnarray}
\par
As is seen from the last equation, the concentration variation in the superfluid liquid mixture is completely due to 
dependence of superfluid density $\rho _s$  upon the $^3$He concentration. 
Since the condition for the thermodynamic stability of liquid mixture implies always $\partial Z/ \partial c >0$ and the superfluid density reduces with the growth  of concentration, i.e. $\partial\rho _s / \partial c <0$, the $^3$He concentration enhances  always in the regions with the larger magnitudes of superfluid velocity. From the physical point of view it is also evident from a possibility to reduce the kinetic energy of the superfluid component $\rho _sV_s^2/2$ with  decreasing the superfluid density. The pressure reduction with the growth of superfluid velocity can be  ascribed to the role of the Bernoulli pressure.
\par
In a dilute unsaturated liquid mixture we have 
 $$
\frac{\partial Z}{\partial c}\approx\frac{T}{m_3c}.
$$ 
This gives the known \cite{Reut} Boltzmann concentration distribution $c(r) =c\,\exp (-U_{\text{eff}}/T)$  with the effective attractive potential   
$$
U_{\text{eff}}(r)=\,\frac{\partial\rho _s/\partial c}{\rho}\,\cdot\,\frac{m_3V_s^2(r)}{2}.
$$
This distribution can only be justified for the distances provided that the $^3$He concentration does not exceed the magnitudes corresponding to the condition of nondegeneracy 
 $$ 
T_F\sim\,\,\frac{\hbar ^2}{2m_3}\left(\frac{\rho c}{m_3}\right) ^{2/3}\,\ll T.
$$ 
\par
For the smaller distances or in the degenerate liquid mixture, the growth of concentration occurs slower and power-like. For the more realistic estimate of the concentration behavior at the close distances to the vortex core, it is necessary to know the detailed behavior of quantities  $Z=Z(P,\, c,\, V_s^2)$ and $\rho _s=\rho _s(P,\,c,\, V_s^2)$.
\par
Note that the $^3$He concentration in the \textit{d}-phase, in any case, cannot exceed magnitude $c_{\lambda}$ corresponding  to the $\lambda$-line at which the superfluid component density $\rho _s$ vanishes or the magnitude $c_{\text{sp}}$  corresponding to the spinodal line at which the derivative $\partial Z/\partial c=0$ and the \textit{d}-phase becomes absolutely unstable. These conditions restrict the minimum size of vortex core. Emphasize that the concentration is $c_{\lambda}>c_{\text{sp}}$ at the temperatures below the tricritical point temperature $T_t$ and, on the contrary, at $T>T_t$ the $\lambda$-line lies ahead of the spinodal and the transition to the \textit{c}-phase occurs continuously in concentration.  Accordingly, we can expect the distinctions in the vortex structure at temperatures above and below the tricritical point temperature, namely, the concentration discontinuity at a vortex radius if $T<T_t$ and the continuous variation  with a kink at the vortex radius if $T>T_t$.
\par
For the temperature region $T>T_t$, when the phase separation of liquid $^3$He-$^4$He mixture does not take place but only the \textit{d}-to-\textit{c} phase transition occurs at some temperature and pressure-dependent concentration $c_{\lambda}$,  the vortex core radius can grow infinitely as the liquid mixture concentration $c$ approaches the $c_{\lambda}$ one.  In fact, for radius $R_{\lambda}$ where the $^3$He impurity concentration reaches  the magnitude $c_{\lambda}$, we have 
$c(R_{\lambda}) = c_{\lambda}(P(R_{\lambda}))$. 
According to Eq. (\ref{f905})
 $$
c_{\lambda}\left(P-\frac{\rho _sV_s^2}{2},\, V_s^2\right) = c -\,\frac{\partial\rho _s/\partial
c}{\rho\,\partial Z/\partial c}\,\frac{V_s^2}{2}
 $$ 
 or approximately on the account of the $\lambda$-line dependence on the pressure and superfluid velocity, we arrive at 
 $$
 c_{\lambda}(P)-\rho
_s\,\frac{dc_{\lambda}}{dP}\,\frac{V_s^2}{2}-\,\frac{\partial\rho _s/\partial V_s^2}{\partial\rho _s/\partial
c}\,V_s^2 = c -\,\frac{\partial\rho _s/\partial c}{\rho\,\partial Z/\partial c}\,\frac{V_s^2}{2}.
 $$ 
Hence we find 
 $$ 
 R_{\lambda}=\frac{m_4\left(-\frac{\partial\rho _s/\partial
c}{\rho\,\partial Z/\partial c}+\,\rho _s\,\frac{dc_{\lambda}}{dP}+2\,\frac{\partial\rho _s/\partial
V_s^2}{\partial\rho _s/\partial c}\right) ^{1/2}}{2\sqrt{2}\,\hbar(c_{\lambda}-c)^{1/2} }\longrightarrow\infty
$$
as $ c\rightarrow c_{\lambda}$. 
\par 
In order to realize the vortex of large core radius below the tricritical point temperature $T_t$, one should have the impurity concentration close to that at the spinodal. In other words, the liquid $^3$He-$^4$He should already be supersaturated, i.e. it should be in the metastable region. As will be seen below, the presence of a vortex, playing a role of nucleation site, prevents from achieving the highly supersaturated state of the \textit{d}-phase. Thus, the vortex core radius remains finite and cannot exceed some maximum value, starting from which the vortex becomes absolutely unstable against the core expansion. This entails the phase separation of a liquid mixture. 
\par
In the \textit{d}-phase the above distinctions and specific features in the vortex core behavior as a function of temperature and concentration could be observed with the aid of second sound. The second sound absorption is very sensitive to the presence of the \textit{c}-phase, allowing us to detect the variation in the total volume of the normal \textit{c}-phase in vortices under varying the liquid mixture concentration. 

\subsection{Energy of nucleation on a vortex}

\par
First, we find the energy of rectilinear vortex in the \textit{d}-phase close to saturation. As a liquid mixture separates, the vortex core is progressively filling with the normal \textit{c}-phase playing a role of a nucleus of new phase. We need to know the energy of the system as a function of vortex core radius $R$. In equilibrium the thermodynamical potentials and temperature  of the normal \textit{c}-phase obey the following requirements: 
\begin{eqnarray*}
\Phi '(P',\, Z',\, T') =\text{const},
\\
Z'(P',\, c',\, T') =\text{const} \;\;\;\text{and}\;\;\;  T'= \text{const}.
\end{eqnarray*}
This yields the ordinary conditions: $c'(r)=\text{const}$ and $P'(r)=\text{const}$.
\par 
The boundary conditions at the vortex core $r=R$, usual for the equilibrium between two phase,  imply an equality of thermodynamical potentials, temperatures, and pressures involving the Laplace pressure: 
\begin{gather}
\Phi '(P', Z', T')  =  \Phi (P, Z, T, V_s)+\frac{V_s^2}{2}  \nonumber 
\\
\equiv  \Phi (P, Z, T,V_s=0)_{|_{r=\infty}}, \nonumber
\\
Z'(P', c', T')  = Z(P, c, T,V_s)  \equiv  Z(P, c, T, V_s=0)_{|_{r=\infty}}, \nonumber
\\
T'  =  T  \;\;\;\text{and}\;\;\;    P'  =  P(R)+\alpha /R . \label{f906}
\end{gather}
šThe surface tension coefficient $\alpha =\alpha (P'-P,\, c'-c)$, in general, depends on a difference of the pressures and concentrations at the interface. For brevity, we omit the notation for temperature. 
\par
Let us denote the impurity concentration far from the vortex at  $r=\infty$ as $c=c_{\text{ps}}(P)+\Delta c$ and, correspondingly, impurity concentration inside the core as  $c'=c'_{\text{ps}}(P)+\Delta c'$. Here $c_{\text{ps}}(P)$ and 
$c'_{\text{ps}}(P)$ are the $^3$He impurity concentrations at the phase separation line. Assuming $\Delta c \ll
c_{\text{ps}}(P)$ and $\Delta c' \ll c'_{\text{ps}}(P)$ as well as $\delta P=\alpha /R+P(r=R)-P$  to be small, we expand 
Eqs.~(\ref{f906}) in deviations 
\begin{gather*}
\Phi '(P,\, c'_{\text{ps}}(P)) +\frac{1+\beta '}{\rho '}\,\Delta P -c'\,\frac{\partial Z'}{\partial c'}\Delta
c' 
\\ 
 =  \Phi (P,\, c_{\text{ps}}(P))-c\,\frac{\partial Z}{\partial c}\Delta c,
\end{gather*}
and
\begin{gather*}
Z'(P,\, c'_{\text{ps}}(P)) -\frac{\beta '}{\rho 'c'}\,\Delta P +\frac{\partial Z'}{\partial c'}\Delta c'  
\\
=Z(P,\, c_{\text{ps}}(P))+\frac{\partial Z}{\partial c}\Delta c  ,
\end{gather*}
where coefficients $\beta$ and $\beta '$ are defined as usual 
 $$
 \beta =\frac{c}{\rho}\,\frac{\partial\rho}{\partial c}\;\;\;\;\text{and}\;\;\;\;\beta ' =\frac{c'}{\rho '
}\,\frac{\partial\rho '}{\partial c'}. 
 $$ 
Hence we arrive readily at the deviations of concentration and pressure from their equilibrium magnitudes inside the vortex  
\begin{gather*}
\Delta P = \rho '\biggl( c'_{\text{ps}}(P)-c_{\text{ps}}(P)\biggr)\frac{\partial Z}{\partial c}\Delta c, 
\\
\Delta c' = \left[1+\beta
'\left(1-\,\frac{c_{\text{ps}}(P)}{c'_{\text{ps}}(P)}\right)\right]\,\frac{\partial Z/\partial c}{\partial
Z'/\partial c'}\Delta c .
\end{gather*}
Thus,  if  \textit{d}-phase is saturated ($\Delta c=0$), the impurity concentration inside the core is equilibrium  $c'=c'_{\text{ps}}(P)$. Introducing the imbalance between the phases 
 $$
\Delta\Phi = [c'_{\text{ps}}(P)-c_{\text{ps}}(P)]\,\frac{\partial Z}{\partial c}\Delta c ,
 $$
we arrive at the equation determining the core radius $R$: 
 $$
\Delta P\equiv \alpha /R +P(R)-P=\rho '\Delta\Phi .
 $$
\par
To find pressure $P(R)$, we involve the behavior of superfluid density  $\rho _s$ as a function of $P$, $c$, $V_s^2$  and use relations~(\ref{f904}). Then we have 
\begin{eqnarray*}
\nabla\rho _s  = \frac{\partial\rho _s}{\partial P}\,\nabla P +\,\frac{\partial\rho _s}{\partial c}\,\nabla
c +\,\frac{\partial\rho _s}{\partial V_s^2}\,\nabla V_s^2 =-\rho_s B\nabla\frac{V_s^2}{2},
\end{eqnarray*}
where $B$ is equal to 
 $$
B=\frac{\rho _s}{\rho}\,\frac{(\partial\log\rho _s/\partial c)^2}{\partial Z/\partial c}+\,
\frac{\partial\rho _s}{\partial P}- 2\,\frac{\partial\log\rho _s}{\partial V_s^2},
 $$
and the superfluid density is approximately given with the following equation: 
 $$
 \rho _s(r)=\rho _s \left(1-B\,\frac{V_s^2(r)}{2}\right). 
 $$
From $\nabla P=-\rho _s(r)\nabla (V_s^2/2)$ we find the pressure 
$$
P(r)=P-\rho _s\frac{V_s^2(r)}{2}+\rho _sB\frac{V_s^4(r)}{8}+\ldots 
$$
\par
Accordingly, core radius $R$ satisfies 
\begin{gather*}
-\rho '\Delta\Phi +\frac{\alpha}{R} -\rho _s\frac{V_s^2(R)}{2}+\rho _s B\frac{V_s^4(R)}{8}+\ldots =0 
\end{gather*}
or
\begin{gather*}
-\rho '\Delta\Phi +\frac{\alpha}{R} -\rho _s\frac{\hbar ^2}{2m_4^2R^2}
+\rho _s B\frac{\hbar ^4}{8m_4^4R^4}+\ldots =0 .
\end{gather*}
Multiplying it with $2\pi R$ and integrating over $R$, we obtain the energy $U(R)$ per unit length for a \textit{c}-phase nucleus in the form of quantized vortex  
\begin{eqnarray}\label{f910}
U(R)=-\rho '\Delta\Phi\,\pi\! R^2 +2\pi\alpha R +\rho _s\frac{\pi\hbar ^2}{m_4^2}\ln \frac{L}{R}\nonumber
\\ 
+\frac{\partial\alpha}{\partial c}\,\frac{\partial\rho _s/\partial c}{\rho\partial Z/\partial c}\,
\frac{\pi\hbar ^2}{m_4^2}\,\frac{1}{R}
-\,\frac{\pi\hbar ^4}{m_4^4}\rho _s\frac{B}{8}\,\frac{1}{R^2}+\ldots
\end{eqnarray}
\par
Here surface tension $\alpha$ is assumed to be only dependent  on the $^3$He impurity concentration at the \textit{c}-\textit{d} interface but a possible pressure and core radius dependence is neglected. The length $L$ is a usual cutoff one for a vortex. The origin for the terms in the vortex energy is obvious. Note only that the third and fifth terms arise wholly from the kinetic energy of superfluid component  $(1/2)\int\rho _s(r)V_s^2(r)\, d^2r$. In essence, equation (\ref{f910}) represents an expansion of linear energy of a nucleus in its inverse radius \cite{BurmiLT}. 

\subsection{Thermal and quantum nucleation rate on the vortex. The crossover temperature to the quantum regime}

\par
We consider here the quantum nucleation rate of \textit{c}-phase on the vortex at zero temperature  and the thermal-quantum crossover temperature in nucleation. First of all, it is necessary to understand the behavior of potential nucleus energy $U(R)$. To simplify and to obtain the clear analytical formulae, we restrict ourselves with the three first terms of expansion in equation  (\ref{f910})
 $$
U(R)=-\rho '\Delta\Phi\,\pi\! R^2 +2\pi\alpha R +\rho _s\frac{\pi\hbar ^2}{m_4^2}\ln \frac{L}{R}\, .
 $$
%%%%%%%%
\widetext
\onecolumngrid
\begin{figure}[h]
\begin{center}
\includegraphics[scale=0.45]{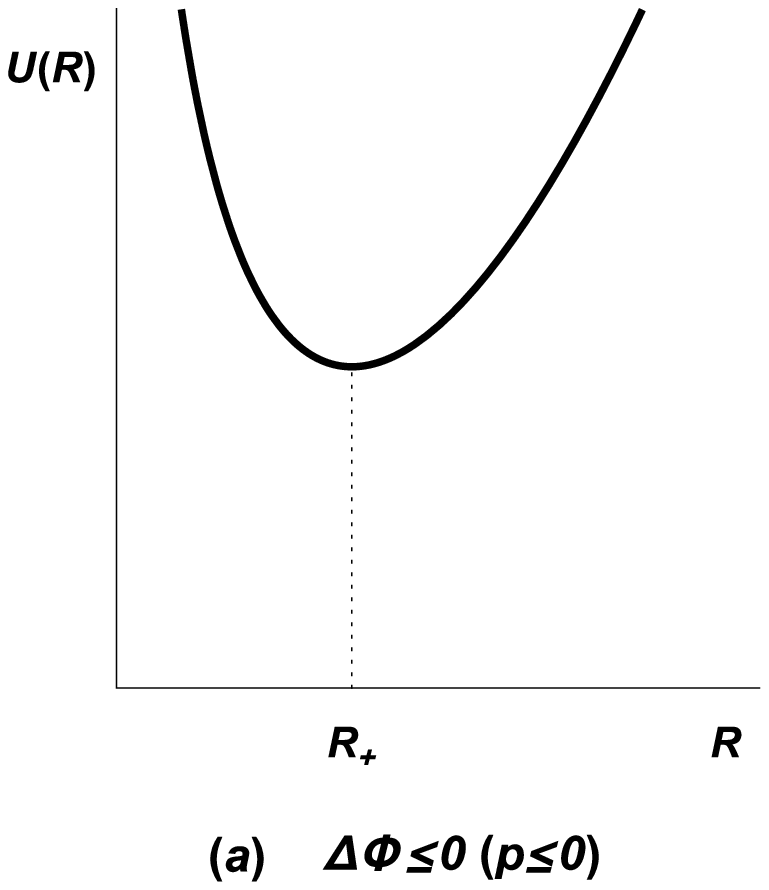}
\includegraphics[scale=0.45]{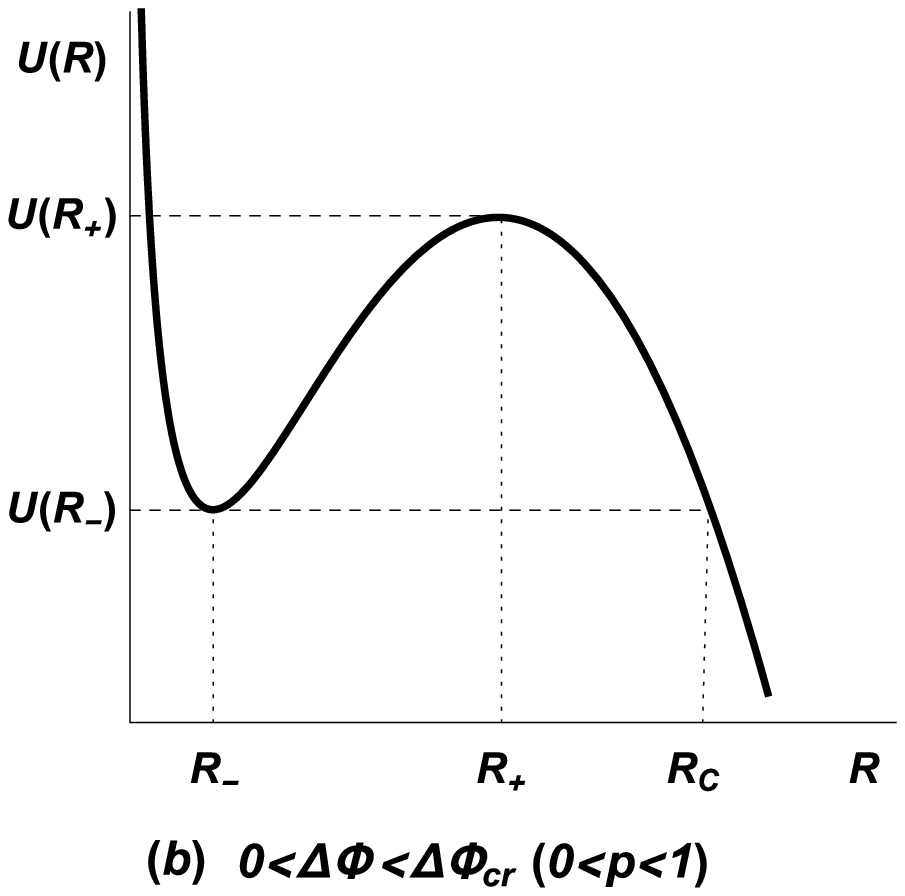}
\includegraphics[scale=0.45]{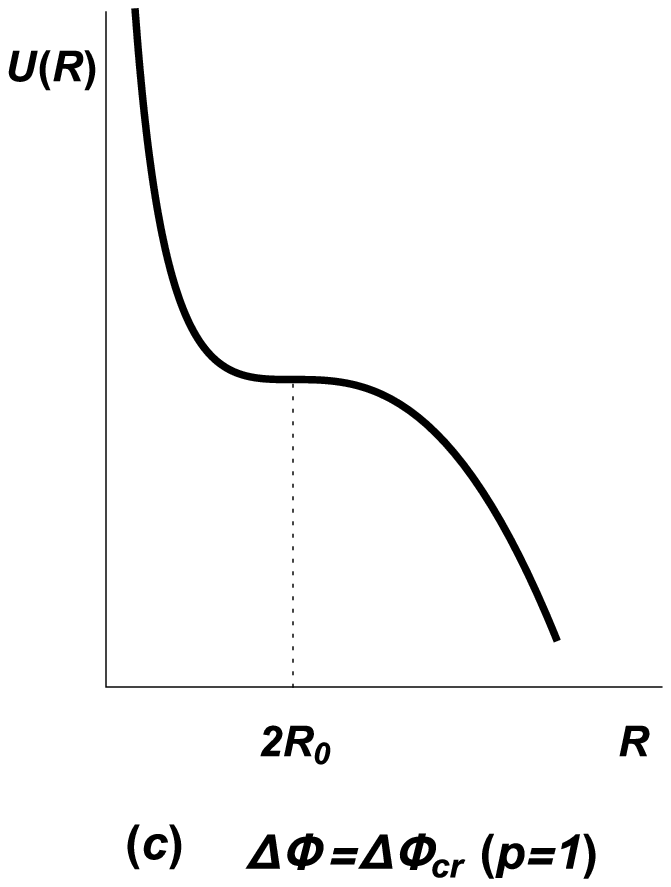}
\includegraphics[scale=0.45]{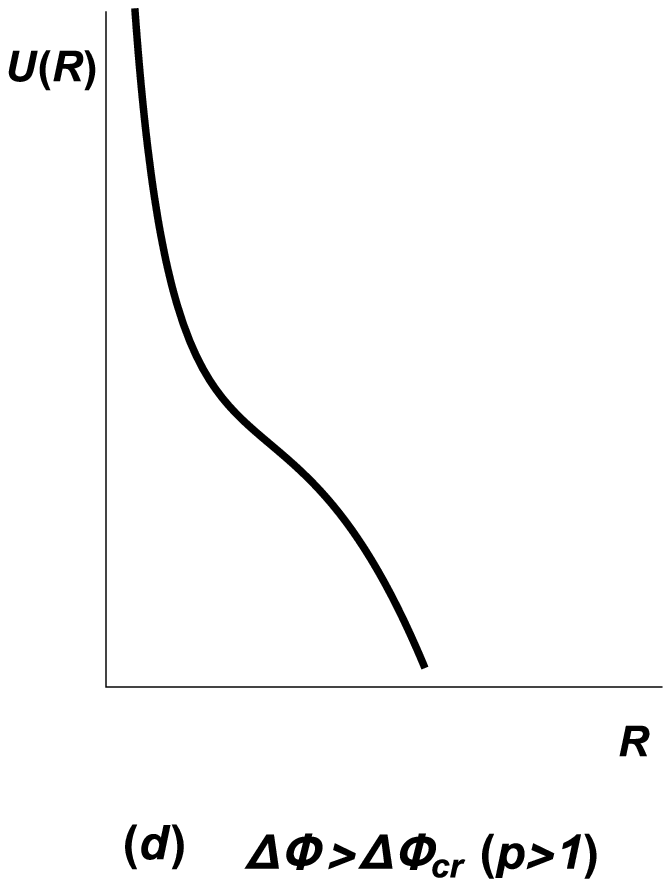}
\caption{The behavior of potential energy $U$ as a function of vortex core radius $R$ in the various ranges of imbalance
 $\Delta\Phi$.} \label{fig91}
\end{center}
\end{figure}
\widetext
\twocolumngrid
%%%%%%%%%%
As is seen from Fig.~\ref{fig91},  energy $U(R)$ as a function of imbalance $\Delta\Phi$ has a various behavior and a different number of extrema depending on the vortex core radius. For unsaturated ($\Delta\Phi <0$) liquid mixture, there is a single extremum  and the corresponding vortex state is absolutely stable (Fig.~\ref{fig91}a). Within the intermediate imbalance range  $0< \Delta\Phi <\Delta\Phi _{\text{cr}}$ there are two extrema in the potential energy (Fig.~\ref{fig91}b) as a function of core radius at 
\begin{gather}
\frac{R_{\pm}}{R_0} = \frac{2}{p}\left(1\pm\sqrt{1-p}\right),\nonumber
\\
R_0 = \frac{\hbar ^2\rho _s}{2\alpha m_4^2}\; , \;\;\;\;\; p=\frac{\Delta\Phi}{\Delta\Phi _{\text{cr}}}\, . 
\label{f912}
\end{gather}
Here $R_0$ is the core radius in the saturated liquid mixture $\Delta\Phi =0$. For the imbalance larger than the critical one  
\begin{equation}\label{f913}
\rho '\Delta\Phi _{\text{cr}}=\frac{\alpha}{4R_0}=\frac{\alpha ^2m_4^2}{2\rho _s\hbar ^2}\, ,
\end{equation}
there are no extrema (Fig.~\ref{fig91}d). This entails an appearance of the line of absolute instability. In other words, vortices which sizes exceed $R_{\text{cr}}=2R_0$ become unstable against vortex core expansion and the phase separation of a liquid mixture proves to be unavoidable. Hence, only for the imbalance range 
$0< \Delta\Phi <\Delta\Phi _{\text{cr}}$  corresponding to $R_0<R<2R_0$, we observe the metastable state which can be destabilized as a result of thermal or quantum fluctuations depending on the temperature.  These specific features are well evident in Fig.~\ref{fig92}. 
%%%%%%%%%%%%
\begin{figure}[h]
\begin{center}
\includegraphics[scale=0.5]{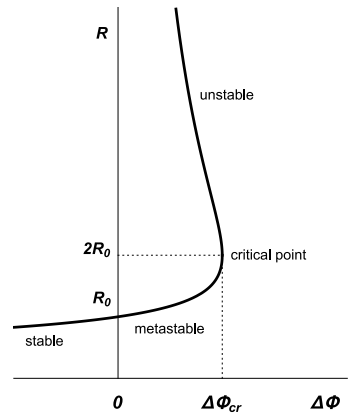}
\caption{The diagram for the equilibrium between vortex line and superfluid $^3$He-$^4$He liquid mixture.  
Here $R$ is the vortex core radius and $\Delta\Phi$ is the imbalance of a liquid mixture. }
\label{fig92}
\end{center}
\end{figure}
%%%%%%%%%%%
\par
Such behavior of nucleus energy as a function of nucleus size and imbalance differs in kind from the case of homogeneous nucleation of spherical drop. The point is that there is a competition of two opposite factors in the presence of a defect similar to vortex. If, for instance, a nucleus grows, the contribution associated with the surface tension increases and the other due to effect of a defect decreases. For negative and small positive  $\Delta\Phi$,  these two contributions result in the minimum for the potential energy (Figs.~\ref{fig91}a and \ref{fig91}b).  On the contrary, if the imbalance is large and the critical nucleus is small, the effect of surface tension is small as well. 
Then, for $\Delta\Phi >\Delta\Phi _{\text{cr}}$, the total energy is determined, in the first turn, with the terms decreasing gradually and, therefore, has neither minimum nor maximum. This results in the unstable state. All these features, involving the similar phase diagrams $R$---$\Delta\Phi$ hold for other defects \cite{Burmi12}, e.g. for a charged ion which influence decays with the distance together with electric field. 
\par
At the first sight, due to existence of critical value $\Delta\Phi _{\text{cr}}$ one may expect that the phase separation of a liquid mixture with the quantized vortex will occur at the same magnitude of imbalance in all experiments. However, the process of phase separation can take place in the metastable region  $0< \Delta\Phi <\Delta\Phi _{\text{cr}}$ before the critical imbalance is achieved. Since the phase separation of metastable state is a random process described with some probability and dependent on imbalance, the experimental magnitudes acquire some dispersion around certain magnitude 
$ \Delta\Phi <\Delta\Phi _{\text{cr}}$.  This magnitude and dispersion of data depend both on the decay probability and on the rate of varying the liquid mixture imbalance in experiment. 
\par
For the high temperatures, the nucleation rate, determined as a nucleation probability per unit time at one nucleation site, is governed with the usual Arrhenius formula for thermal fluctuations 
 $$
\Gamma _{\text{cl}}=\nu\exp (-U_L/T)
 $$
where $\nu$ is the frequency of attempts. The activation energy $U_L$ for a vortex of length $L$ is expressed as 
$U_L = L\,\Delta U_0$ with $\Delta U_0=U(R_{+})-U(R_{-})$ determined from the difference between the maximum and the minimum of energy $U(R)$ (Fig.~\ref{fig92}b):  
\begin{eqnarray*}
\frac{\Delta U_0}{2\pi\alpha R_0}= \frac{2}{p}\sqrt{1-p}-\ln\frac{1+\sqrt{1-p}}{1-\sqrt{1-p}}
\\ 
=\left\{
\begin{array}{ccc}
\frac{4}{3}(1-p)^{3/2} & \text{for} & p\rightarrow 1, 
\\
\\
\frac{2}{p}-\ln\frac{4}{p} & \text{for} & p\rightarrow 0.
\end{array}
\right.
\end{eqnarray*}
It is seen that the nucleation rate enhances drastically as $p=\Delta\Phi /\Delta\Phi _{\text{cr}}\rightarrow 1$ due to vanishing the potential barrier. 
\par
As the temperature approaches absolute zero temperature, the quantum fluctuations become predominant. To estimate the quantum nucleation rate, we employ the theory of quantum nucleation in the two-dimensional systems \cite{Burmi94}. Within the exponential accuracy we estimate the zero-temperature nucleation rate as 
 $$
\Gamma _{\text{q}}=\nu _{\text{q}}\exp (-A_L/\hbar ). 
 $$
Here $\nu _{\text{q}}$ is the attempt frequency and $A_L=AL$ is the doubled underbarrier action for vortex length $L$ where $A$ is the quantity per unit length of a vortex. 
\par
To calculate the quantum probability, it is necessary to estimate the effective mass of the expanding vortex core. We treat the boundary conditions at the vortex core surface $r=R(t)$ as for the boundary between two different phases. First of all, we have a conservation for the radial component of mass flow across the boundary   
\begin{gather*}
-\rho '\dot{R}  =  \jmath _r(R)-\rho\dot{R},
\\
\bm{\jmath} =  \rho _n\bm{V}_n + \rho _s\bm{V}_s . 
\end{gather*}
Here we assume that the \textit{c}-phase of density $\rho '$ in the vortex core is at rest ($V'=0$). The second condition is a continuity for the tangential component of momentum flux density tensor  $\Pi _{ik}$
$$
\Pi '_{\theta\, r}- \jmath\, ^{\prime}_{\theta}\dot{R} =  \Pi _{\theta\, r}-\jmath _{\theta}\dot{R},
$$
where for the superfluid phase 
$$
\Pi _{ik}=\rho _n V_{n\, i}V_{n\, k}+\rho _{s\, i}V_{n\, i}V_{s\, k}+P\delta _{ik}. 
$$
Accordingly, we have 
$$
\rho _n V_{n\,\theta}(V_{n\, r}-\dot{R})+\rho _s V_{s\,\theta}(V_{s\, r}-\dot{R})=0. 
$$
Treating the normal component of superfluid \textit{d}-phase as a liquid with the properties of an ordinary viscid fluid, we should require an equality of tangential components for the velocities of adjacent fluids. Then we have 
$$
V_{n\,\theta}(R)=V'_{\theta}(R)=0. 
$$
From the above three equations we find the magnitudes of velocities at the boundary $r=R(t)$
\begin{gather*}
V_{s\, r}(R)  =  \dot{R}, 
\\
V_{n\, r}(R)  =  -\,\frac{\rho '-\rho _n}{\rho _n}\,\dot{R},
\\
V_{n\,\theta}(R)  =  0. 
\end{gather*}
šAs is seen, the mass transfer across the surface of the expanding vortex core is connected with the normal component flow alone. 
\par
On the neglect of the compressibity of a liquid mixture the distribution for the radial components of superfluid and normal velocities is equal to 
\begin{gather*}
V_{s\, r}(r) =  V_{s\, r}(R)R/r,
\\
V_{n\, r}(r)  = V_{n\, r}(R)R/r\; , \;\;\;\; r>R(t). 
\end{gather*}
In the logarithmic quasistationary approximation and on the analogy with the two-dimensional nucleation \cite{Burmi94} we can estimate the kinetic energy of the expanding vortex core as 
$$
2\pi\rho _{\text{eff}}R^2\ln\frac{\Lambda (R)}{R}\,\frac{\dot{R}^2}{2}=\frac{1}{2}\, M(R)\dot{R}^2
$$
with the following effective density $\rho _{\text{eff}}$
\begin{equation}\label{f921}
\rho _{\text{eff}}=\rho _s+\frac{(\rho '-\rho _n)^2}{\rho _n}.
\end{equation}
The cutoff parameter $\Lambda (R)$ is of the order of the sound velocity multiplied with the typical time of the core expansion 
$\Lambda (R)\approx s_{\text{eff}}\,\tau (R)$. 
Since there are two sound velocities in the superfluid liquid mixture, velocity $s_{\text{eff}}$ is some weighted average of 
first and second sound velocities.  In other words, length $\Lambda (R)$ is an effective size of sound propagation region for the time of the underbarrier nucleation evolution, i.e.,  size of the perturbed medium surrounding the nucleus. Here we do not consider possible energy dissipation effects due to viscosity and diffusion in the process of vortex core growth.  
\par
So, action $A$ is calculated between the classical turning points in the potential $\Delta U(R) = U(R)-U(R_{-})$ as 
\begin{equation}\label{f921a}
 A=2\int _{R_{-}}^{R_c}\sqrt{2M(R)\Delta U(R)}\, dR
 \end{equation}
where $R_c$ is the exit point from the barrier  $U(R_c)=U(R_{-})$. Since the logarithm is a slowly varying function, for our aim it is sufficient to estimate the growth time of a nucleus as follows: 
$$ 
\tau \approx \left(\frac{M(R)R^2}{2U(R)}\right)^{1/2}_{R\approx R_c}. 
$$
\par
The analytical expressions for the effective action are succeeded to obtain for the two limiting cases.  For the small degree of  imbalance $p=\Delta\Phi /\Delta\Phi _{\text{cr}}\ll 1$, the contribution to the vortex energy, associated with the superfluid motion, plays a minor role and the effective action is mainly governed with the magnitude of the surface  tension:   
\begin{gather*}
A(T=0) =\frac{32\sqrt{2}\pi ^2}{p^{5/2}}\left(\alpha\rho _{\text{eff}}\, R_0^5\ln\frac{s^2_{\text{ef}}R_0}{\alpha
p}\right) ^{1/2}
\\ 
\propto (\Delta\Phi)^{-5/2}\ln ^{1/2}(\Delta\Phi )^{-1}.
\end{gather*}
The quantum critical radius $R_c$ exceeds significantly the core radius $R_0$ in a weakly supersatured liquid mixture 
 $$
R_c = \frac{2\alpha}{\rho '\Delta\Phi}=\frac{8R_0}{p}\gg R_0. 
 $$
The typical time of nucleus growth reads 
$$ 
\tau (R_c)\approx\left(\frac{\rho _{\text{eff}}R_c^3}{8\alpha}\right) ^{1/2}\propto (\Delta\Phi )^{-3/2}.  
$$
\par
Let us turn to the other limit $p=\Delta\Phi /\Delta\Phi _{\text{cr}}\rightarrow 1$. In this case the imbalance is close to the line of absolute instability when the potential barrier, separating two states, vanishes. In this case, as usual, the potential barrier can 
be approximated with a cubic parabola 
$$ 
\frac{\Delta U(R)}{2\pi\alpha R_0}=\sqrt{1-p}\left(\frac{R-R_{-}}{R_{-}}\right) ^2 -\,
\frac{1}{3}\left(\frac{R-R_{-}}{R_{-}}\right) ^3.
$$
As $p\rightarrow 1$, the distance between the points of entrance $R_{-}$ and exit $R_c$ reduces to 
$$ 
R_c-R_{-}=3R_{-}\,\sqrt{1-p} 
$$
but the growth time of a nucleus increases  
$$ 
\tau\approx \left(\frac{8\rho _{\text{eff}}R_0^3}{\alpha\sqrt{1-p}}\right) ^{1/2}. 
$$
With the help of Eq.~(\ref{f921a}) the action per unit length can be estimated as 
 $$
A=\frac{192\pi}{5}(1-p)^{5/4}\left(\alpha\rho _{\text{eff}}\,R_0^5\ln\frac{8s_{\text{eff}}^2\,\rho
_{\text{eff}}\, R_0}{\alpha\sqrt{1-p}}\right) ^{1/2}\! . 
 $$
Thus,  effective action vanishes with approaching to the instability. Accordingly, the probability of quantum nucleation grows drastically.  
\par
The next important point in the low temperature kinetics is a crossover temperature $T_q$ between the quantum and classical nucleation regimes. A simple estimate for $T_q$ can be obtained from comparing the classical and quantum actions under assumption $\nu _q\approx \nu$. So, 
$$
T_q =\hbar U_L/A_L .
 $$
Hence we have in the limit of small $p=\Delta\Phi /\Delta\Phi _{\text{cr}}\ll 1$ imbalance 
\begin{gather}\label{f925}
T_q=\frac{p^{3/2}}{8\sqrt{2}\pi }\biggl(\frac{\alpha\hbar ^2}{\rho _{\text{eff}}\, R_0^3\ln [s^2_{\text{eff}}\,\rho
_{\text{eff}}\, R_0/(\alpha p)]}\biggr) ^{1/2}\! .
\end{gather}
In this region of imbalance the crossover temperature grows as the imbalance enhances. 
\par
On the other hand, as the imbalance tends to the critical value, i.e.  $1-p\ll 1$,  the crossover temperature starts to reduce according to relation 
\begin{gather}\label{f925a}
T_q=\frac{5}{72}(1-p)^{1/4}\biggl(\frac{\alpha\hbar ^2}{\rho _{\text{eff}}\, R_0^3\ln [8s^2_{\text{eff}}\,\rho
_{\text{eff}}\, R_0/(\alpha \sqrt{1-p})]}\biggr) ^{1/2}\!  . 
\end{gather}
%%%%%%%%%%
\begin{figure}[h]
\begin{center}
\includegraphics[scale=0.5]{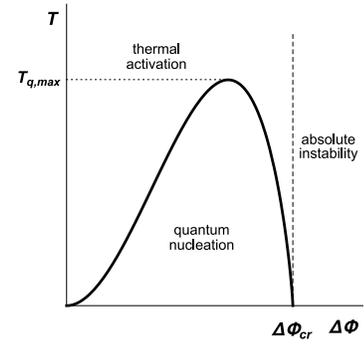}
\caption{The various types of the \textit{c}-phase nucleation onto a quantized vortex in a supersaturated $^3$He-$^4$He liquid mixture. The solid line shows the thermal-quantum crossover temperature.  The dashed line is the spinodal, separating the metastable states from absolutely unstable ones.} \label{fig93}
\end{center}
\end{figure}
%%%%%%%%%
\par 
The total behavior for the crossover temperature $T_q$ as a function of imbalance is shown in Fig.~\ref{fig93}. Note here that the crossover temperature maximum  $T_{q,\, \text{max}}$ is shifted in the direction to the line of absolute instability. Another specific feature, apparently, inherent in all transitions near the spinodal, is that the nucleation mechanism becomes again the thermal one instead of quantum as a function of supersaturation $\Delta\Phi$ or imbalance degree \cite{Burmi12} in the intermediate vicinity to the line of absolute instability (spinodal). This takes place though the temperature is less than $T_{q,\,\text{max}}$.  For a rectilinear vortex, the crossover temperature $T_q$ is naturally independent of its length. 
\par
We do not consider here the effect of dissipative phenomena on the nucleation rate of \textit{c}-phase nucleus at the quantized vortex core, which can be associated with viscosity, impurity $^3$He diffusion, and impossibility of using the quasistationary approximation for rectilinear vortex. In principle, one can here distinguish the hydrodynamical and ballistic regimes of nucleus growth. However, for the region of supersaturation close to critical value, a possibility  of hydrodynamical $R_{c} \gg l(T)$ regime, where $l(T)$ is the mean free path of excitations in a liquid mixture, is unlikely in the quantum nucleation region since the quantum vortex core radius  $R_c$ in the nucleus growth does not exceed several vortex core radii  $R_0$ in the saturated liquid mixture (about a few tens of angstrom). For the ballistic  $R_{c}\ll  l(T)$ regime, one can suppose that the friction coefficient is directly proportional to the core area. Accordingly, the friction coefficient per unit length is $\mu (R)\propto R$. 
In the viscous $M(R)\tau ^{-1}(R)\ll \mu (R)$  limit a simple estimate gives 
\begin{gather*}
T_q \approx  \frac{\alpha\hbar}{2R_0\,\mu(2R_0)}\sqrt{1-\frac{\Delta\Phi}{\Delta\Phi _{\text{cr}}}}  ,  \;\;\;
 \Delta\Phi\rightarrow\Delta\Phi _{\text{cr}}\, ;
 \\
A(T<T_q) \approx  2\pi\biggl( 1-\frac{\Delta\Phi}{\Delta\Phi _{\text{cr}}}\biggr) (2R_0)^2\mu (2R_0)\biggl(
1-\frac{T^2}{3T_q^2}\biggr)  
\end{gather*}
and in the other  $\Delta\Phi\ll\Delta\Phi _{\text{cr}}$   limit
\begin{gather*}
T_q  \approx   \frac{2\pi\alpha\hbar}{\mu (R_c)R_c}\propto (\Delta\Phi)^2 ,   \;\;\; R_c=2R_0, \;\;\;\;
(\Delta\Phi\ll\Delta\Phi _{\text{cr}})\, ;
\\
A(T<T_q)\approx  \mu (R_c)R_c^2\propto (\Delta\Phi )^{-3}  .
\end{gather*}
\par 
Comparing these formulae with the previous ones, it is seen that the qualitative character for the behavior of effective action $A(T)$ and crossover temperature $T_q$  remains unchanged as a function of imbalance $\Delta\Phi$. The diagram 
$T_q$--$\Delta\Phi$ conserves its shape in kind (Fig.~\ref{fig93}) though the quantum nucleation region reduces and the thermal activation region increases beside the instability line. 

\subsection{Rapid nucleation line and the numerical estimates.}

\par
Below we discuss some consequences from the equations above and perform the numerical estimates connected with the 
\textit{c}-phase nucleation on a quantized vortex in the supersaturated superfluid $^3$He-$^4$He liquid mixture. First, we analyze in kind the possible positions of the rapid nucleation line in the $T$--$\Delta\Phi$ diagram of nucleation regimes. The rapid nucleation line exists also as in the case of homogeneous nucleation of spherical drops \cite{LPK,BDT} due to very drastic dependence of nucleation rate on the imbalance. The rapid nucleation line separates the region where the nucleation rate is practically zero and supersaturated liquid mixture does not separate infinitely long on the time scale of experimental period 
from the region where the phase separation occurs almost instantaneously. 
\par
After preparing the metastable state $0< \Delta\Phi < \Delta\Phi _{\text{cr}}$ at temperature $T$ the liquid mixture separates 
eventually for the expectation time $\tau _{obs}$. Thus the nucleation probability is approximately equal to unity\begin{equation}\label{f926}
W(\Delta\Phi ,\, T,\, \tau _{obs}, \, N_{nuc})\equiv \tau _{obs}\, N_{nuc}\,\Gamma (\Delta\Phi ,\, T)=1. 
\end{equation}
Here $\Gamma$ stands for either $\Gamma _{cl}$ or $\Gamma _q$ in the correspondence with the temperature range and  $N_{nuc}$ is the number of nucleation sites. Equation (\ref{f926}) determines the rapid nucleation line $\Delta\Phi _{\text{sat}}(T)$ in diagram $T$--$\Delta\Phi$. This corresponds to the experimentally achievable supersaturation.  
\par
For the nucleation probability, we have 
\begin{eqnarray*}
W=\tau _{obs}N_{nuc}\nu e^{-A(\Delta\Phi , T)/\hbar}\hspace{2cm}
\\
=\left\{
\begin{array}{ll}
\tau _{obs}N_{nuc}\nu _{cl}\exp (-U_L(\Delta\Phi\, , T) \;\; \text{if} \;\;\; T> T_q(\Delta\Phi ), 
\\
\tau _{obs} N_{nuc}\nu _q\exp (-A_L(\Delta\Phi\, , T) \;\; \text{if} \;\;\; T< T_q(\Delta\Phi ).
\end{array}
\right.
\end{eqnarray*}
Hence one can see that the position of the rapid nucleation line depends on the temperature, the number of nucleation sites, and the rate of sweeping the liquid mixture imbalance. Since the shape of potential energy $U(R)$ depends strongly on the closeness to the instability line, the effect of the sweep rate on the position of the rapid nucleation line here is more essential as compared with the case of homogeneous nucleation.  
\par
\begin{figure}[h]
\begin{center}
\includegraphics[scale=0.45]{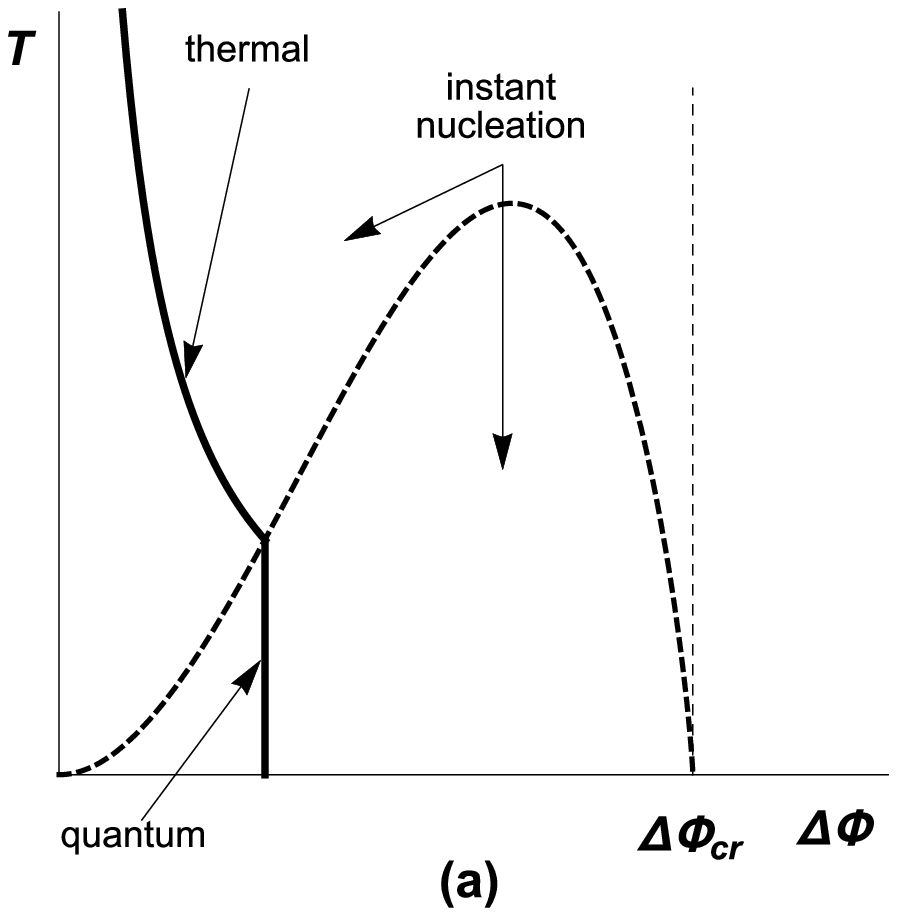}
\includegraphics[scale=0.45]{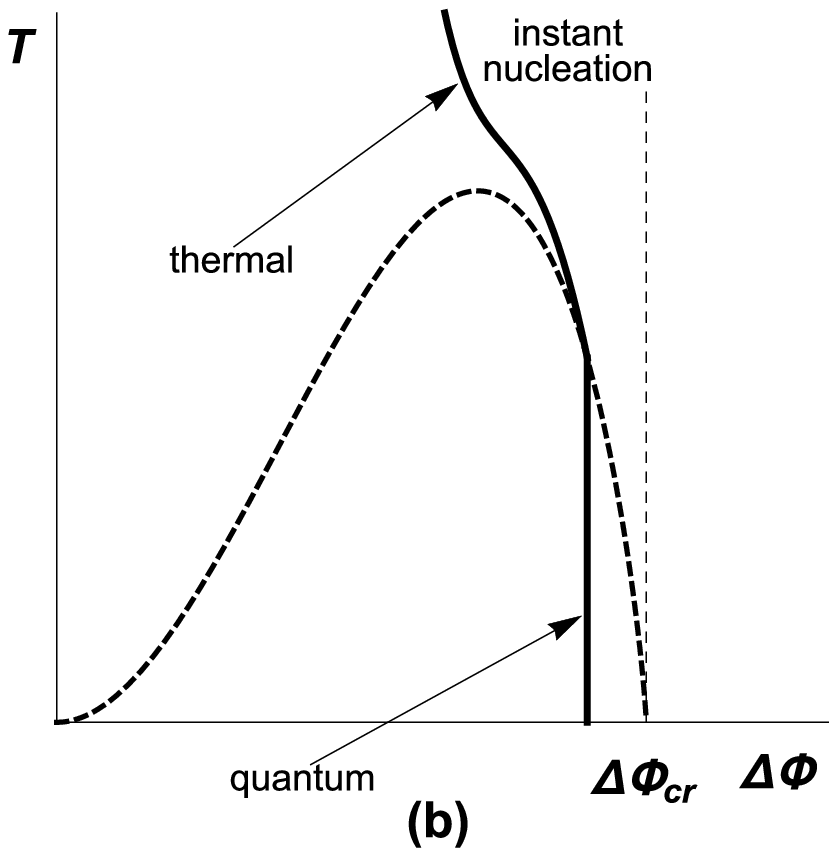}
\caption{The schematic for the rapid nucleation lines (solid lines): (a) for low nucleation rates $\Gamma$ and large expectation time $\tau _{obs}$, (b) for high nucleation rates $\Gamma$ and small expectation time $\tau _{obs}$. }\label{fig4}
\end{center}
\end{figure}

Depending on the expectation time $\tau _{obs}$ and the number of nucleation sites, one can discern two opposite cases in the position of the rapid nucleation line in the $T$--$\Delta\Phi$ diagram 
(Fig.~\ref{fig4}).  The first case is restricted with the inequality  
\begin{equation}\label{f927}
\ln (\nu\tau _{obs}N_{nuc})\gg 32\sqrt{2}\pi ^2\biggl(\frac{\alpha\rho _{\text{eff}}R_0^5L^2}{\hbar ^2}
\ln\frac{s^2_{\text{eff}}\rho _{\text{eff}}R_0}{\alpha}\biggr) ^{\! 1/2}
\end{equation}
and corresponds to the limit of low nucleation rates $\Gamma$. This corresponds to the large lifetime of a supersaturated liquid mixture against the decay channel considered. In this case  (Fig.~\ref{fig4}a) the rapid nucleation line lies far from the instability line and $\Delta\Phi _{\text{sat}} \ll \Delta\Phi _{\text{cr}}$. Therefore, the existence of the instability line has no significant effect on the nucleation kinetics. In the classical thermal activation region the attainable supersaturation is strongly temperature-dependent according to $\Delta\Phi _{\text{sat}}\propto 1/T$. In the quantum $T<T_q$ region the attainable supersaturation is almost independent of temperature. Correspondingly, the crossover temperature  $T_q$, proportional to 
$(\Delta\Phi _{\text{sat}}) ^{3/2}$, is significantly smaller than the maximum crossover temperature $T_{q,\, \text{max}}$. 
%%%%%%%%
%%%%%%%
\par
For the opposite case of high nucleation rates when inequality (\ref{f927}) is invalid, the existence of instability affects essentially the position of the rapid nucleation line at sufficiently low temperatures  (Fig.~\ref{fig4}b). As the temperature lowers, the rapid nucleation line should closer approach the instability line since the smallness of potential barrier can compensate a decrease of temperature in the classical exponent, providing us the high nucleation rates. As a result, in the thermal activation regime the temperature behavior for the attainable critical supersaturations should go over from drastic  
$\Delta\Phi _{\text{sat}}\propto 1/T$ to the smoother one 
$$ 
\Delta\Phi _{\text{sat}}=\Delta\Phi _{\text{cr}}\left(1-(T/T_*)^{2/3}\right) 
$$
in the low temperature region if $T\lesssim T_{q,\, {\text{max}}}$. Here $T_*$ is some typical temperature which can be determined from Eq.~(\ref{f926}) with the classical exponent at $p\rightarrow 1$. From the experimental point of view this 
distinctive feature, associated with the closeness to instability, can deliver some trouble in determining the crossover temperature between the classical and quantum regimes, imitating the genuine crossover with the transition to almost temperature-independent behavior for the observable imbalance of a liquid mixture.
\par
Another specificity is associated with the presence of two regions for the thermal activation regime at various $\Delta\Phi$ for the same temperature  $T< T_{q,\, {\text{max}}}$ (Fig.~\ref{fig93}).  However, as is seen from Fig.~\ref{fig93}a and  \ref{fig93}b, the observation of such reentrant behavior is impossible under the fixed nucleation rate $\Gamma$. To do this, it is necessary to vary any parameter in equation (\ref{f926}), e.g. the number of nucleation sites $N_{nuc}$.  In liquid $^3$He-$^4$He mixture this can be done with introducing vortices intentionally before the start of the phase separation process in the \textit{d}-phase of a liquid mixture. 
\par
Let us turn to the numerical estimates of the results obtained. We start from the calculation of the critical value  
$\Delta\Phi _{\text{cr}}$ which plays a key role in comprehending the phase separation in a superfluid liquid mixture with quantized vortices. As for the surface tension, we take the magnitude for the flat interface  between the bulk phases of liquid mixture at zero pressure $\alpha =$0.0239 erg/cm$^2$ \cite{Balfour}. Using simplest estimates  (\ref{f912})
and (\ref{f913}), we find $\rho '\Delta\Phi _{\text{cr}}\approx$0.011~bar and, correspondingly,  $R_{\text{cr}}=$12.7~\AA . 
A relatively large critical vortex core justifies an applicability of macroscopic approximation to some extent. On account of estimating the derivative $\partial Z/\partial c\approx$0.9$\cdot$10$^8$~erg/g for the limiting critical value of saturation responsible for the vortex core instability, we arrive at  $\Delta x _{\text{cr}}=$(1.9--2.0)\% \cite{BurmiLT}. Thus the estimate agrees with that obtained in work~\cite{Jezek95}. 
\par
Expansion (\ref{f910}) in the inverse core radius involves the inhomogeneous distribution of concentration, pressure and superfluid density in the \textit{d}-phase. However, the effect of these terms proves to be small and counts about 5\%. 
If we neglect the contribution associated with coefficient $B$ in  (\ref{f910}), putting $B=0$, and expecting the order-of-magnitude estimate of $\partial\alpha /\partial c\approx\alpha$, we find somewhat smaller but the close value for the limiting supersaturation  $\Delta x_{\text{cr}}=$1.80\%.  This is connected with the relatively large critical core radius. Note also that value $R_{\text{cr}}=$12.7~\AA\,  correlates with the core radius $R_{\text{sp}}\approx$12.5~\AA\, calculated under assumption that the $^3$He concentration at the core boundary corresponds to the spinodal of the bulk \textit{d}-phase and equals $x_{\text{sp}}\sim$16\%. 
\par
The important characteristic for nucleation kinetics is the thermal-quantum crossover temperature $T_q$.  To estimate the latter,  it is necessary to know the effective density $\rho _{\text{eff}}$ (\ref{f921}) which proves to be approximately equal to 
$\rho _{\text{eff}}\approx 2.2\rho _s$ at the phase separation line and lies within the range 0.23 -- 0.40 g/cm$^3$.  The highest possible crossover temperature $T_{q,\, \text{max}}$ can be estimated either from (\ref{f925}) at $p=1$ or from  (\ref{f925a}) at $p=0.5$. This results in relatively small temperatures $T\leqslant T_{q,\, \text{max}}\lesssim$2.3~mK necessary for a possible observation of quantum nucleation.  
\par
To conclude, we have analyzed the \textit{c}-phase nucleation in the supersaturated \textit{d}-phase when a quantized vortex plays a role of nucleation site. Though the maximum possible supersaturation of the \textit{d}-phase in the presence of vortices proves to be in the almost satisfactory agreement with the observable magnitudes of supersaturation, the estimate for the crossover temperature to the quantum nucleation regime is rather small as compared with the temperature at which the 
transition in temperature behavior is observed for the critical supersaturation $\Delta x_{\text{cr}}$ of a liquid mixture. 
Varying the values of physical parameters in order to increase $T_{q,\, \text{max}}$, we have an enhancement of the  
maximum attainable concentration at which the vortex core instability occurs. Then the agreement with experiment becomes worse in this parameter. So, the assumption about the heterogeneous phase-separation mechanism in a supersaturated superfluid $^3$He-$^4$He liquid mixture with quantized vortices as nucleation sites, apparently,  cannot describe the experimentally observed picture of phase separation on the whole. 
\par
An obstacle for quantitative comparison arises from the exponential behavior of nucleation rate. In such situation from the experimental point of view it would be useful to study the effect of the number of nucleation sites on the phase separation of a liquid mixture under the planned and controlled introduction of quantized vortices into the \textit{d}-phase.  One of 
possibilities is an experiment in a rotating cryostat and the study of the phase separation rate  of liquid mixture  as a function of rotation velocity. Since in the limit of small density of  vortices the nucleation rate is proportional to the number of vortex lines, the observable nucleation rate should also be proportional to the rotation velocity. 
\par 
For the large rotation velocities, especially when the spacing between the vortex lines is comparable with the core sizes, the critical value of supersaturation $\Delta\Phi _{\text{cr}}$ and the potential barrier, separating the metastable state from unstable,  are strongly suppressed. Accordingly, the phase separation rate of a liquid mixture should drastically grow in the limit of high rotation velocities. In addition, due to difference in the centrifugal energy of $^3$He and $^4$He atoms the spatial $^3$He distribution becomes inhomogeneous over the bulk of a rotating fluid, facilitating the reduction of the critical supersaturation 
$\Delta\Phi _{\text{cr}}$. In any case it is known that the $^3$He impurities strongly affect the process  of nucleating the quantized vortices in superfluid $^4$He.

\end{document}